\documentclass[aps,prl,10pt,twocolumn,superscriptaddress,longbibliography,nobalancelastpage]{revtex4-1}

\usepackage{amsmath}
\usepackage{amssymb}
\usepackage{graphicx}
\usepackage{dsfont}
\usepackage{braket}
\usepackage{bm}
\usepackage{wrapfig}
\usepackage[caption=false]{subfig}
\usepackage{kantlipsum}
\usepackage{mathtools}
\usepackage[colorlinks=true,linkcolor=red,urlcolor=magenta,citecolor=blue]{hyperref}
\usepackage{tikz}
\usepackage[english]{babel}

\newcommand{\paper}[0]{paper}

\definecolor{g}{rgb}{.1,0.4,.1} 
\definecolor{b}{rgb}{0,0.2,1}
\definecolor{rouge}{rgb}{0.82,0.,0.}
\definecolor{vert}{rgb}{0.,0.82,0.}
\definecolor{orange}{rgb}{1,0.5,0.}
\definecolor{bleu}{rgb}{0.,0.,0.82}
\definecolor{m}{rgb}{0.82,0.,0.82}
\definecolor{vert2}{rgb}{0.,0.5,0.}
\definecolor{rougeclair}{rgb}{1.0,0.7,0.7}
\definecolor{gris}{rgb}{.8,.8,.8} 

\newcommand{\one}{\mathds{1}}

\newcommand{\drawone}{\draw[line width=1pt]}
\newcommand{\drawhalf}{\draw[line width=1.5pt]}
\newcommand{\drawtensor}[2]{
\drawone (#1-1,#2+1) -- (#1+2,#2+1); \drawone (#1+2,#2+1) -- (#1+1,#2-1); 
\drawone (#1+1,#2-1) -- (#1-2,#2-1); \drawone (#1-2,#2-1) -- (#1-1,#2+1);
\drawone(#1+0,#2+0) -- (#1+0,#2+2); 
\drawone (#1+.5,#2+1) -- (#1+1,#2+2); \drawone (#1-.5,#2-1) -- (#1-1,#2-2); 
\drawone (#1+1.5,#2+0) -- (#1+3,#2+0); \drawone (#1-1.5,#2+0) -- (#1-3,#2+0);}
\newcommand{\drawdot}[2]{\draw[fill] (#1,#2) circle (.2);}
\newcommand{\drawred}{\draw[line width=1pt, color=red]}

\newcommand{\tikzTNO}[4]{
\begin{tikzpicture}[baseline = (X.base),every node/.style={scale=1},scale=.30]
\drawone (-1,1) -- (2,1); \drawone (2,1) -- (1,-1); \drawone (1,-1) -- (-2,-1); \drawone (-2,-1) -- (-1,1);
\drawone (0,0) -- (0,2); \drawone(0,-2) -- (0,-1);
\drawone (.5,1) -- (1,2); \drawone (-.5,-1) -- (-1,-2); \drawone (1.5,0) -- (2.5,0); \drawone (-1.5,0) -- (-2.5,0); 
\draw (0,0) node (X) {$\phantom{X}$};
\draw (-3,0) node {$#1$}; \draw (3,0) node {$#3$};
\draw (1.25,2.5) node {$#4$}; \draw (-1.25,-2.5) node {$#2$};
\end{tikzpicture}}

\newcommand{\tikzTNOt}[4]{
\begin{tikzpicture}[baseline = (X.base),every node/.style={scale=1},scale=.50]
\drawhalf (-1,1) -- (2,1); \drawhalf (2,1) -- (1,-1); \drawhalf (1,-1) -- (-2,-1); \drawhalf (-2,-1) -- (-1,1);
\drawhalf (1.25,.5) -- (1.25,2); \drawhalf (-1.25,-.5) -- (-1.25,1); 
\drawhalf (-0.625,.5) -- (-0.625,2); \drawhalf (+0.625,-.5) -- (+0.625,1);
\drawhalf (1.25,-.5) -- (1.25,-1); \drawhalf (-1.25,-1) -- (-1.25,-2);
\drawhalf (+0.625,-1) -- (+0.625,-2);
\drawhalf (.5,1) -- (1,2); \drawhalf (-.5,-1) -- (-1,-2); \drawhalf (1.5,0) -- (2.5,0); \drawhalf (-1.5,0) -- (-2.5,0); 
\draw (0,0) node (X) {$\phantom{X}$};
\draw (-3,0) node {$#1$}; \draw (3,0) node {$#3$};
\draw (1.25,2.5) node {$#4$}; \draw (-1.25,-2.5) node {$#2$};
\draw (1,.5) node {$b$}; \draw (-1,-.5) node {$c$};
\draw (-.375,.5) node {$a$}; \draw (.375,-.5) node {$d$};
\end{tikzpicture}}

\begin{document}

\title{Bridging Perturbative Expansions with Tensor Networks}

\author{Laurens Vanderstraeten}
\email{laurens.vanderstraeten@ugent.be}
\affiliation{Department of Physics and Astronomy, Ghent University, Krijgslaan 281, S9, 9000 Gent, Belgium}

\author{Micha\"el Mari\"en}
\affiliation{Department of Physics and Astronomy, Ghent University, Krijgslaan 281, S9, 9000 Gent, Belgium}

\author{Jutho Haegeman}
\affiliation{Department of Physics and Astronomy, Ghent University, Krijgslaan 281, S9, 9000 Gent, Belgium}

\author{Norbert Schuch}
\affiliation{Max-Planck-Institut f\"{u}r Quantenoptik, Hans-Kopfermann-Str. 1, 85748 Garching, Germany}

\author{Julien Vidal}
\affiliation{Laboratoire de Physique Th\'{e}orique de la Mati\`{e}re Condens\'{e}e, CNRS UMR 7600, Universit\'{e} Pierre et Marie Curie, 4 Place Jussieu, 75252 Paris Cedex 05, France}

\author{Frank Verstraete}
\affiliation{Department of Physics and Astronomy, Ghent University, Krijgslaan 281, S9, 9000 Gent, Belgium}
\affiliation{Vienna Center for Quantum Science and Technology, Faculty of Physics, University of Vienna, Boltzmanngasse 5, 1090 Vienna, Austria}

\begin{abstract}
We demonstrate that perturbative expansions for quantum many-body systems can be rephrased in terms of tensor networks, thereby providing a natural framework for interpolating perturbative expansions across a quantum phase transition. This approach leads to classes of tensor-network states parametrized by few parameters with a clear physical meaning, while still providing excellent variational energies. We also demonstrate how to construct perturbative expansions of the entanglement Hamiltonian, whose eigenvalues form the entanglement spectrum, and how the tensor-network approach gives rise to order parameters for topological phase transitions.
\end{abstract}

\maketitle

In the last decade, interest in two-dimensional strongly correlated quantum systems has increased considerably, mainly due to the existence of exotic quantum behavior and topological order. The strong quantum correlations and entanglement patterns that characterize these systems give rise to, e.g., topologically protected ground states \cite{Wen90_2,Bravyi10} and anyonic excitations \cite{Wilczek82, Kitaev06}, which can be used to manufacture reliable quantum memories and perform fault-tolerant quantum computation \cite{Freedman03, Ogburn99, Wang_book}. However, the entanglement properties that allow for these nontrivial properties also make these systems very hard to simulate, as mean-field approaches fail to capture the essential quantum correlations and quantum Monte Carlo simulations often suffer from the sign problem.

Perturbation theory has proven successful in studying the robustness of topological phases (see Ref.~\cite{Dusuel11} for instance), but such an approach necessarily fails when approaching a transition point. Indeed, quantum phase transitions naturally disconnect different perturbative descriptions, so that the critical properties cannot be captured easily. A complementary approach is the variational method, which does not inherently break down at criticality. In that respect, much progress has been obtained using tensor-network states \cite{Verstraete08,Orus14}, although it remains a matter of debate to what extent variational calculations correctly describe the phase of a given model \cite{Balents14}.

This \paper{} proposes a method that combines the perturbative and variational approaches. The central idea is to represent the perturbative expansions of the ground state in the tensor-network formalism, lift the perturbative coefficients to variational parameters, and apply the tensor-network machinery to variationally optimize the energy density. This approach allows to merge distinct perturbative expansions in a single variational wave function, and to bridge between perturbative series expansions on both sides of a critical point.

\noindent\emph{General framework}---
Let us start with standard perturbation theory for a Hamiltonian $H=\lambda_0 \, H_0+\lambda_1 \,H_1$. At order one, in the limit where $\lambda_i \gg \lambda_j$, the (unnormalized) perturbed ground state of $H$ is given by 
\begin{equation}
\ket{\psi} = \left(\mathds{1}- \frac{\lambda_j}{\lambda_i} \, V_i^j \right) \ket{\psi_i},\label{eq:standardpert}
\end{equation}
where \mbox{$V_i^j=\displaystyle{\frac{\mathds{1}-P_i}{H_i-E_i} H_j}$} and $P_i=\ket{\psi_i}\bra{\psi_i}$. For many-body systems, such expressions do not have the extensivity structure expected for a ground-state wave function. If $H_j$ is a sum of local interactions, the perturbative correction creates a zero-momentum superposition of local excitations rather than a finite density of excitations on top of the unperturbed reference state $|\psi_i\rangle$. The exponentiated form $\mathrm{e}^{- (\lambda_j/\lambda_i) \, V_i^j } \ket{\psi_i}$, instead, does give rise to an extensive wave function by automatically incorporating all disconnected Feynman diagrams. An elegant way of obtaining such an expression is via the formalism of quasiadiabatic time evolution \cite{Hastings05}, which shows that perturbation theory can be cast into a low-depth quantum circuit acting on a reference state \cite{VanAcoleyen16}.

Starting from the two unperturbed states $\ket{\psi_0}$ and $\ket{\psi_1}$, we can construct a one-parameter path $\ket{\psi_\alpha}$ interpolating between these two states, connect both perturbative expansions into a single wave function
\begin{equation}
\ket{\beta_1,\beta_0,\alpha} = \exp\left(- \beta_1 \, V_0^1 \right)\exp\left(- \beta_0 \, V_1^0 \right) \ket{\psi_\alpha},
\label{eq:pert_exp}
\end{equation}
and consider $(\alpha, \beta_0,\beta_1)$ as variational parameters. The variational optimization requires us to compute the energy expectation value accurately, a task that might seem more easy using the original expression of Eq.~\eqref{eq:standardpert}. However, because of the nonextensivity, Eq.~\eqref{eq:standardpert} generically yields a zero variational contribution, a dilemma pointed out by Feynman \cite{Feynman87}. Here, we rely on the low-depth quantum circuit representation of Eq.~\eqref{eq:pert_exp} and use tensor-network methods to compute the corresponding expectation values. In fact, the formalism of tensor networks \cite{Verstraete08} provides a direct way to build extensive wave functions that match, order by order, a given perturbative expansion. Indeed, according to the linked-cluster theorem, in order to represent increasing orders in perturbation theory, we need to apply clusters of local operators of increasing size on our reference state. These clusters can be efficiently encoded in a tensor-network operator \cite{Pirvu10}, so that, if the reference state $\ket{\psi_\alpha}$ itself can be represented as a tensor-network state, we can construct states similar to the one in Eq.~(\ref{eq:pert_exp}) in an efficient way.

\par In two dimensions this construction gives rise to projected entangled-pair states \cite{Verstraete04} (PEPSs), for which efficient algorithms exist \cite{Vanderstraeten16} to optimize the variational parameters directly in the thermodynamic limit. The computational complexity of the tensor-network construction is determined by the so-called bond dimension $D$, which scales linearly in the number of clusters (this number typically scales exponentially in the order of perturbation theory). Because we only have a small number of parameters, the variational optimization can be performed with high precision and does not introduce systematic errors.
\par Additionally, rephrasing perturbation theory in terms of tensor networks gives rise to a description of correlations in terms of an auxilliary space encoding the entanglement degrees of freedom. This allows us to determine a perturbative expansion of the entanglement Hamiltonian \cite{Cirac11}, whose eigenvalues represent the entanglement \mbox{spectrum \cite{Li08}}.

%
%
\noindent\emph{Transverse-field Ising model}---
%
%
Let us illustrate our approach with the ferromagnetic transverse-field Ising model (TFIM) on a square lattice defined by 
%
\begin{equation*}
H_{\text{TFIM}} =-\lambda_0 \sum_i X_i -\lambda_1 \sum_{\langle ij \rangle}Z_iZ_j.
\end{equation*}
%
%
Here, $X_i$ and $Z_i$ denote the usual Pauli matrices acting on site $i$.  This model is known to exhibit a phase transition between a polarized phase and a symmetry-broken phase detected by the magnetization along the $Z$ direction. The critical point is located at \mbox{$\lambda_0/ \lambda_1 = 3.04438(2)$} as estimated by Monte Carlo simulations \cite{Blote02}. Let us note that the ground-state mani\-fold of $H_1$ is twofold degenerate but, since $\bra{-Z} H_0^n \ket{+Z}=0$ for any finite $n$ in the thermodynamic limit, we can safely use the procedure described above (in the following, $\ket{\pm A}$ denotes the polarized state in the $\pm A$ direction).

For this problem, we choose the one-parameter reference state
%
%
\begin{equation}
\ket{\psi_\alpha} = \prod_{i} (\mathds{1}+\alpha Z_i) \ket{+X},
\label{eq:ising0}
\end{equation}
%
%
which is a simple product state interpolating between the unique ground state $ \ket{\psi_{0}} = \ket{+X} $ at $\lambda_1=0$ and one of the symmetry-broken ground states $\ket{\psi_{1}}=\ket{+Z}$ at $\lambda_0=0$. If one only considers $\ket{\psi_\alpha}$ as an order-zero variational ansatz, one finds a critical point at \mbox{$\lambda_0/ \lambda_1 = 4$}, which is the standard mean-field result.

\begin{table} \centering
\begin{tabular}{|p{0.47\columnwidth}|p{0.47\columnwidth}|}
\hline
$ \;\tikzTNO{1}{0}{0}{0}\rightarrow \sqrt{\frac{\lambda_1}{4\lambda_0}} Z \; $  &
$ \; \tikzTNO{1}{1}{0}{0}\rightarrow \frac{\lambda_1}{2\lambda_0} \one \; $ \\
\hline
&
$ \; \tikzTNO{1}{0}{1}{0}\rightarrow \frac{\lambda_1}{2\lambda_0} \one \; $ \\
\hline
\end{tabular}
\caption{The tensor network operator reproducing perturbation theory up to second order around the polarized state $\ket{+X}$. The left (right) column represents the first- (second-) order terms. Other entries are obtained by symmetry.} \label{tab:pert}
\end{table}

Next, we consider the first- and second-order perturbative expansions in terms of tensor-network operators acting on $|\psi_0\rangle$ and $|\psi_1\rangle$. The wave function corresponding to those expansions can be represented as a PEPS with bond dimension $D=2$. In the limit $\lambda_0\gg\lambda_1$, we can reproduce the first-order perturbative wave function by a simple tensor-network operator. When acting on the unperturbed state $\ket{\psi_0}$, it gives rise to a PEPS with bond dimension $D=2$. Although not exactly equal to the exponentiated operator, this state reproduces the correct perturbative wave function up to first order. Because of its extensivity, the operator also creates disconnected pairs of $Z_iZ_j$ clusters in second order, but we need to introduce two new entries in our tensor network in order to account for the next-nearest-neighbour clusters $\sum_{\langle\braket{ij}\rangle} Z_i Z_j$. This results in the $D=2$ tensor network operator listed in Table~\ref{tab:pert}, which can reproduce the perturbative expansion up to second order when acting on $\ket{\psi_0}$. A similar construction can be done starting from the ferromagnetic state $|\psi_1 \rangle$, where second-order perturbation theory yields $\sum_{\braket{ij}} X_i X_j$.

\begin{table} \centering
\begin{tabular}{|p{0.47\columnwidth}|p{0.47\columnwidth}|}
\hline
$ \;\tikzTNO{0}{0}{0}{0} \rightarrow \one + \beta_0 X \; $ &
$ \; \tikzTNO{1}{0}{0}{0} \rightarrow \gamma_0 X \; $ \\
\hline
$ \;\tikzTNO{2}{0}{0}{0}\rightarrow \beta_1 Z \; $  &
$ \; \tikzTNO{2}{2}{0}{0}\rightarrow \gamma_1 \one \; $ \\
\hline
&
$ \; \tikzTNO{2}{0}{2}{0}\rightarrow \gamma_1 \one \; $ \\
\hline
\end{tabular}
\caption{The tensor-network operator for the TFIM containing four variational parameters. The above entries are all nonzero entries of the tensor (up to rotations) out of which the tensor-network operator is built. The left (right) column contains entries needed to reproduce the first- (second-) order expansion.} \label{tab:TNOtable}
\end{table}

A key feature is that both of those tensor networks with $D=2$ can be implemented as a single tensor network with bond dimension $D=3$. The perturbative coefficients can then be lifted to variational parameters to provide an ansatz wave function with five variational parameters, two ($\beta_0,\beta_1$) corresponding to the first-order expansion, two ($\gamma_0,\gamma_1$) for second order and one ($\alpha$) for the reference state $\ket{\psi_\alpha}$ on which the tensor-network operator acts (see Table \ref{tab:TNOtable}).

%
%
\begin{figure}[t]
\includegraphics[width=\columnwidth]{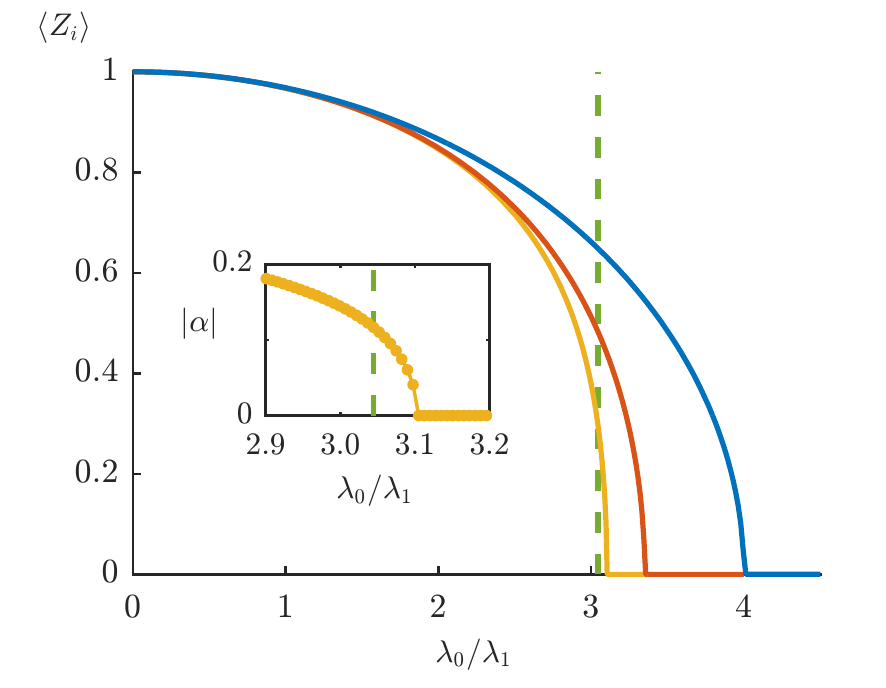}
\caption{Magnetization along the $Z$ direction in the TFIM as a function of the magnetic field for different variational ans\"atze:  order zero (blue) with one parameter ($\alpha$); order one (red) with three parameters ($\alpha,\beta_0,\beta_1$);  order two (yellow) five parameters  ($\alpha,\beta_0,\beta_1,\gamma_0,\gamma_1$) (see Table \ref{tab:TNOtable} for details). Inset: behavior of the virtual order parameter $\alpha$ near the critical point.}
\label{fig:TFIM}
\end{figure}
%
%

Using this ansatz and the techniques developed in Ref.~\cite{Vanderstraeten16} to minimize the energy, we computed the optimal parameters as a function of $\lambda_0/\lambda_1$. The ground-state magnetization as a function of the magnetic field is shown in Fig.~\ref{fig:TFIM}. The order-one ansatz yields a critical transition at \mbox{$\lambda_0/ \lambda_1 \simeq 3.35$}, whereas one gets \mbox{$\lambda_0/ \lambda_1 \simeq 3.1$} for the order-two ansatz. These results show that we can compute the phase diagram of the TFIM with a good precision by systematically adding quantum fluctuations on top of the mean-field state $\ket{\psi_\alpha}$ as suggested by perturbation theory. Interestingly, the $\mathds{Z}_2$-symmetry breaking is solely determined by $\alpha$ defined in Eq.~\eqref{eq:ising0}, so that this parameter can also be considered as a reliable ``virtual order parameter'' (see inset in Fig.~\ref{fig:TFIM}).

%
%
\noindent\emph{Entanglement Hamiltonians}---
%
Our perturbative PEPS wave function now allows us to define an entanglement Hamiltonian, which we illustrate by considering the perturbative expansion around the polarized state $\ket{+X}$ as defined by the tensors in Table \ref{tab:pert}. The entanglement Hamiltonian is defined as \mbox{$\exp(-H_{\mathrm{ent}})=\sqrt{\rho_L}\rho_R\sqrt{\rho_L}$}, where $\rho_L$ and $\rho_R$ are the leading left and right eigenvectors of the quantum transfer matrix associated with the PEPS \cite{Cirac11}. In the present case, the transfer matrix is real symmetric, and $\rho_L=\rho_R=\rho$ can easily be obtained perturbatively. In second order (see Table \ref{tab:pert}) we obtain a density matrix of a spin-1/2 chain:
\begin{multline}
\rho = \prod_{k} \left( P_k^+ + \frac{\epsilon}{4} P_k^- \right) \\
 + \frac{\epsilon^2}{32} \sum_{\braket{ij}}  \big( 9 \, X_i  X_{j}  - Y_{i}  Y_{j} \big) \prod_{k\neq i,j} P_k^{+} + \mathcal{O}(\epsilon^3), 
\end{multline}
where $P^{\pm }_k=\frac{\one \pm Z_k}{2}$ is the projector onto the polarized state in the direction $\pm Z$ acting on site $k$, and $\epsilon=\lambda_1/\lambda_0$. By using the expansion 
\begin{multline*}
\mathrm{e}^{2A + \epsilon B} = 
\mathrm{e}^{A}\bigg[\one + \epsilon B+\epsilon \sum_{n=1}^{+\infty} \frac{\overbrace{[A,[\ldots,[A}^{2n},B]]]}{(2n+1)!} +\mathcal{O}(\epsilon^2)  \bigg] \mathrm{e}^A,
\end{multline*}
we find that the entanglement Hamiltonian:
\begin{multline}
-H_{\text{ent}} = \log\left(\frac{\epsilon}{4}\right)  \sum_i (\one - Z_i) 
 + \epsilon\sum_{\braket{ij}}\left(X_iX_j+Y_iY_j\right)  \\ 
 -\frac{5\epsilon^2}{8}\log\left(\frac{\epsilon}{4}\right)\sum_{\braket{ij}}\left(X_i X_j-Y_iY_j\right) 
\end{multline}
reproduces the above expression for $\rho=\exp(-H_\text{ent})$ up to second order in $\epsilon$. It is fascinating that we obtain a nearest-neigbor XY spin chain in a transverse magnetic field that reproduces the perturbative expansion for the entanglement spectrum around the polarized state. 

Our framework now also enables a perturbative calculation of the entanglement entropy for e.g.\ a bipartion of an infinitely long cylinder with circumference $L$ into two half-infinite cylinders; we find the result
\begin{equation}
S_{\text{ent}} = L  \frac{\epsilon^2}{16} \left[1- 2\log\left(\frac{\epsilon}{4}\right)\right] + \mathcal{O}(L\epsilon^4\log\epsilon).
\end{equation}

%
%
\noindent\emph{Toric code in a magnetic field}---
%
%
The second example we consider is the toric code \cite{Kitaev03} in a magnetic field. The phase diagram of this model has been the subject of many studies for various directions of the field \cite{Hamma08, Trebst07, Vidal09_1, Vidal09_2, Tupitsyn10, Dusuel11}. Here, for simplicity, we consider the case where the field points in the $X$ direction. This model, called the TCX model in the following, is defined as
%
%
\begin{equation*}
H_{\text{TCX}} =-\lambda_0 \sum_i X_i- \lambda_1\left(\sum_{v} A_v  + \sum_{p} B_p\right).
\end{equation*}
%
%
Degrees of freedom are spin 1/2 living on the links of a square lattice. Vertex and plaquette operators are defined by
%
%
\begin{equation*}
A_v= \prod_{i \in v}X_i, \quad
B_p= \prod_{i \in p}Z_i,
\end{equation*}
%
%
where products are performed over all sites belonging to vertex $v$ and plaquette $p$, respectively. As early realized \cite{Hamma08,Trebst07}, the TCX model can be mapped onto the TFIM if one restricts to the (charge-free) sector where $\braket{A_v}=+1$. It is easy to see that the ground state belongs to this sector for all $\lambda_0$ and $\lambda_1$. For $\lambda_0 \gg \lambda_1$ the system is in a polarized phase, whereas for $\lambda_0 \ll \lambda_1$ the system is in a topologically ordered phase that strongly contrasts with the symmetry-broken phase of the TFIM. The mapping onto the TFIM exchanges $\lambda_1$ and $\lambda_0$ so that the transition point of the TCX is found at \mbox{$\lambda_0/\lambda_1=1/3.04438(2)\simeq 0.328$}.

%
%
\begin{figure} \centering
\includegraphics[width=\columnwidth]{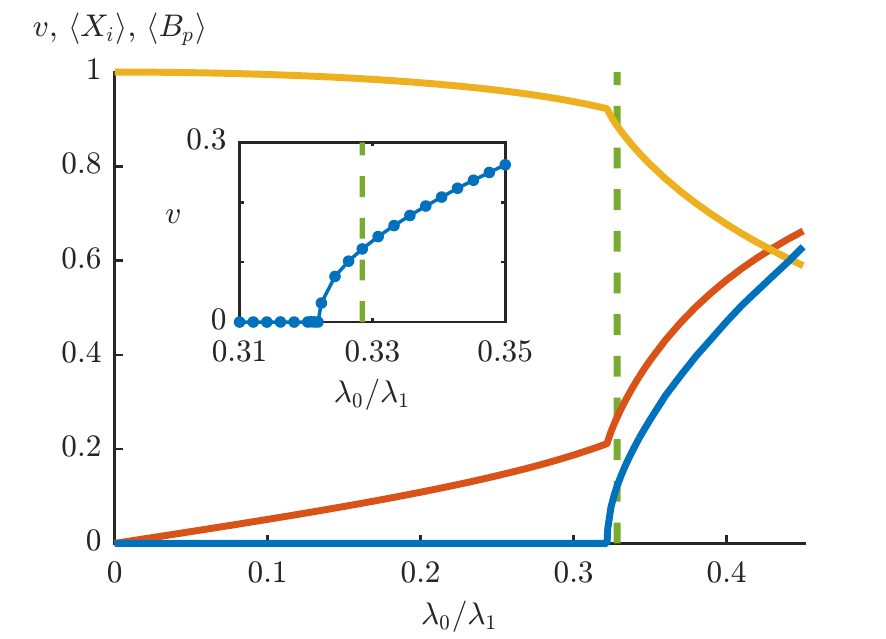}
\caption{Virtual order parameter $v$ (blue), magnetization along the $X$ direction (red), and the expectation value of the operator $B_p$ (yellow) in the TCX model as a function of the magnetic field obtained with the $D=4$-PEPS ansatz (fifteen parameters). The vertical line (dashed green) indicates  the critical point $\lambda_0/\lambda_1 \simeq 0.328$. Inset: behavior of the virtual order parameter $v$ near the critical point.}
\label{fig:TCX}
\end{figure}
%
%

Following Ref.~\cite{Dusuel15}, we choose as a reference state
%
%
\begin{equation}
\ket{\psi_\alpha} = \prod_p \left( \one + \alpha \,B_p \right) \ket{+X}.
\label{eq:toric0}
\end{equation}
%
%
For $\lambda_1=0$, $\ket{\psi_0}$ is the polarized ground state of $H_0$, whereas, for $\lambda_0=0$, $\ket{\psi_1}$ is a ground state (among four under periodic boundary conditions) of $H_1$. Let us note that, contrary to the TFIM, $\ket{\psi_\alpha}$ given in Eq.~(\ref{eq:toric0}) is not a product state (except for $\alpha=0$), but an entangled state described by a PEPS with bond dimension $D=2$ \cite{Verstraete06}. From a variational point of view, both models are thus very different. Using only $\ket{\psi_\alpha}$, one finds a transition for $\lambda_0/\lambda_1=1/4$ \cite{Dusuel15}, in analogy with the TFIM.

We can now repeat the same procedure as before and include quantum fluctuations on top of this mean-field state $\ket{\psi_\alpha}$ as inspired by perturbation theory. For the case of second-order perturbation theory, this leads to a PEPS with bond dimension $D=4$ with fifteen free parameters. As for the TFIM, $\alpha$ defined in Eq.~(\ref{eq:toric0}) can be considered as a virtual order parameter. Indeed, at order zero, it has been shown in Ref.~\cite{Dusuel15} that the topological entanglement entropy \cite{Kitaev06_2, Levin06} is only nonvanishing for $\alpha = 1$. More generally, the concept of $\mathsf{G}$ injectivity \cite{Schuch10} allows us to characterize the topological content of our variational states by identifying the entanglement symmetry that is realized on the virtual level of the tensor-network operator. Away from $\alpha=1$, this virtual symmetry is explicitly broken, but also for $\alpha=1$ the virtual symmetry can be broken spontaneously \cite{Schuch13,Haegeman15,Duivenvoorden17}. Both explicit and spontaneous breaking of the virtual symmetry, and thus of the physical topological order, can be detected by a suitable order parameter evaluated on the virtual level of the tensor network. For the case at hand, we find that the variational solution favors explicit symmetry breaking, which we quantify as $v=\lVert A-u(A)\rVert $, where $u(\cdot)$ represents the nontrivial $\mathbb{Z}_2$ operation on the virtual level \cite{suppmat}. In Fig.~\ref{fig:TCX}, we have plotted this virtual order parameter as a function of $\lambda_0/\lambda_1$, as well as the expectation values of the relevant operators in $H_0$ and $H_1$. We observe that the phase transition is signaled by a sharp onset of the virtual order parameter, and we estimate the transition point at $\lambda_0/\lambda_1\simeq0.322$.

\noindent\emph{Conclusions}--- %
In this \paper{}, we have used insights from perturbation theory to motivate a variational ansatz for the description of ground states of perturbed Hamiltonians in two-dimensional spin systems in the thermodynamic limit. This ansatz is a subset of PEPS, but contains only a handful of parameters related to operators whose presence is expected from perturbation theory. Crucially, we have shown that the ansatz manages to smoothly interpolate between perturbative expansions on either side of a critical point. In addition, our approach gives access to (i) a virtual order parameter, which, in the absence of symmetry breaking, signals a topological phase transition, and (ii) the entanglement spectrum of perturbative wave functions giving rise to an explicit construction of the entanglement Hamiltonian.

Our method is generally applicable to quantum lattice models for which one can write down perturbative expansions. In the case of the square-lattice Heisenberg model, where the first-order perturbative expansion starting from the Ising limit can be written as a PEPS with bond dimension $D=2$, we have observed that this one-parameter family of states gives rise to the same variational energy as a full $D=2$-PEPS simulation for the isotropic Heisenberg point. It should be interesting to improve this result by going to higher orders in perturbation theory, connecting different perturbative expansions, and by considering other two-dimensional lattices. Furthermore, our method is particularly suited to tackle a range of topological phase transitions among which is the perturbed string-net model \cite{Levin05, Schulz13, Schulz14, Dusuel15, Schulz16}.

From the perspective of PEPS simulations our results are important for the following reasons. First of all, whereas the physical meaning of the variational parameters in a PEPS has always been somewhat of a mystery, the parameters in our ansatz have a clear meaning in terms of perturbative expansions. Second, from a numerical perspective such a reduced parametrization has clear advantages. Although the optimization problem is drastically reduced in size, we can still capture the critical behavior across (topological) quantum phase transitions. Also, we open up a way to apply the formalism of virtual PEPS symmetries characterizing topological phases \cite{Schuch10, Sahinoglu14, Bultinck15} to the variational simulation of topological phase transitions. The virtual order parameter that we have identified for the toric code model serves as a first illustration of what is possible in that direction. 

%
\noindent\emph{Acknowledgements}--- %
We acknowledge inspiring discussions with S. Dusuel. This work was supported by the Austrian Science Fund (FWF) through grants ViCoM and FoQuS and the European Commision through SIQS and ERC grants QUTE, WASCOSYS (No. 636201) and ERQUAF (No. 715861). J.H., M.M., and L.V. are supported by the Research Foundation Flanders (FWO).

\bibliography{./bibliography}

\allowdisplaybreaks[1]
\cleardoublepage
\begin{center}
\textbf{\large SUPPLEMENTAL MATERIAL}
\end{center}

In this supplemental material we elaborate on the construction of tensor-network operators (TNOs) that we have used in the main body of the text. We start by writing down the necessary formulas of standard perturbation theory, show how a tensor-network operator (TNO) is constructed, and then explain the TNO constructions for the transverse-field Ising model (TFIM), the toric code in magnetic field (TCX), and the square-lattice XXZ model in some detail.

\section{Perturbation theory}

General perturbation theory treats the Hamiltonian
\begin{equation}
H = H_0 + \lambda V,
\end{equation}
by writing down the wave function and ground-state energy as a series in $\lambda$:
\begin{align}
\ket{\psi} &= \ket{\psi_0} + \lambda \ket{\psi_1} + \lambda^2 \ket{\psi_2}+  \mathcal{O}(\lambda^3),\\
E &= E_0 + \lambda E_1 + \lambda^2 E_2 + \mathcal{O}(\lambda^3).
\end{align}
Expanding the Schr\"{o}dinger equation in different orders gives rise to the following equations
\begin{align}
(H_0-E_0) \ket{\psi_0} &= 0 ,\\
(H_0-E_0) \ket{\psi_1} &= (E_1 - V) \ket{\psi_0}, \\
(H_0-E_0) \ket{\psi_2} &= (E_1-V) \ket{\psi_1}  + E_2 \ket{\psi_0}. \label{eq:pert_second}
\end{align}
Solving the first-order equation gives the result
\begin{align}
E_1 &= \bra{\psi_0}V\ket{\psi_0}, \\
\ket{\psi_1} &= - \frac{\one-P_0}{H_0-E_0} V \ket{\psi_0}.
\end{align}
We observe that $\braket{\psi_0|\psi_1}=0$. Solving the second-order equation gives
\begin{equation}
E_2 = \bra{\psi_0}V\ket{\psi_1},
\end{equation}
and
\begin{multline}
(H_0-E_0) \ket{\psi_2} = E_1 \ket{\psi_1} - (\one-P_0) V \ket{\psi_1} \\ - P_0 V \ket{\psi_1} + E_2 \ket{\psi_0},
\end{multline}
so that
\begin{multline} \label{eq:pert_psisecond}
\ket{\psi_2} = - \left(\frac{\one-P_0}{H_0-E_0}\right)^2 V P_0 V \ket{\psi_0} \\ + \frac{\one-P_0}{H_0-E_0} V \frac{\one-P_0}{H_0-E_0} V \ket{\psi_0},
\end{multline}
because the last two terms cancel.
\par Note that the state $\ket{\psi}$ that we obtain is not normalized, but we can add a component along the unperturbed state to $\ket{\psi_2}$ which makes sure that the state is normalized up to second order in $\lambda$:
\begin{multline}
\ket{\psi_2} = - \left[ \left(\frac{\one-P_0}{H_0-E_0}\right)^2 V P_0 V + \left(\frac{\one-P_0}{H_0-E_0} V \right)^2  \right. \\ \left. -\frac{1}{2} P_0 V \left( \frac{\one-P_0}{H_0-E_0}\right)^2 V \right] \ket{\psi_0}.
\end{multline}
If one is not concerned with the normalization of the state $\ket{\psi}$, we can equally well work with the form in Eq.~\eqref{eq:pert_psisecond}. In fact, one can always add a component along the unperturbed state to the state $\ket{\psi_2}$ without this having an influence on the second-order equation [Eq.~\eqref{eq:pert_second}].

\section{Tensor-network operators}
\label{sec:TNO}

\begin{figure} \centering
\qquad \begin{minipage}{.25\columnwidth}
\subfloat[]{%
$\begin{tikzpicture}[baseline = (X.base),every node/.style={scale=1.3},scale=.4]
\draw (0,-9) node (X) {$\phantom{X}$};
\drawone (-1,-5) -- (2,-5); \drawone (2,-5) -- (1,-7); \drawone (1,-7) -- (-2,-7); \drawone (-2,-7) -- (-1,-5);
\drawone (0,-6) -- (0,-4); \drawone(0,-8) -- (0,-7);
\drawone (.5,-5) -- (1,-4); \drawone (-.5,-7) -- (-1,-8); \drawone (1.5,-6) -- (2.5,-6); \drawone (-1.5,-6) -- (-2.5,-6);
\drawone (0,-7) -- (0,-8); \drawone (-.5,-8) -- (.5,-8); \drawone (-.5,-9) -- (0,-9.5); \drawone (+.5,-9) -- (0,-9.5);
\draw (0,-8.6) node {$\psi$};
\end{tikzpicture} $%
}
\end{minipage}%
\begin{minipage}{.65\columnwidth}
\subfloat[]{%
{$\begin{tikzpicture}[baseline = (X.base),every node/.style={scale=1},scale=.45]
\drawone (-1,1) -- (2,1); \drawone (2,1) -- (1,-1); \drawone (1,-1) -- (-2,-1); \drawone (-2,-1) -- (-1,1);
\drawone (0,0) -- (0,2);
\drawone (.5,1) -- (1,2); \drawone (-.5,-1) -- (-1,-2); \drawone (1.5,0) -- (2.5,0); \drawone (-1.5,0) -- (-2.5,0);
\draw (0,0) node (X) {$\phantom{X}$};
\end{tikzpicture}$}
}\\
\subfloat[]{%
{$\begin{tikzpicture}[baseline = (X.base),every node/.style={scale=1},scale=.20]
\drawtensor{0}{1} \drawtensor{6}{1} \drawtensor{12}{1}
\drawtensor{-2}{-3} \drawtensor{4}{-3} \drawtensor{10}{-3}
\drawtensor{-4}{-7} \drawtensor{2}{-7} \drawtensor{8}{-7}
\end{tikzpicture}$}
}
\end{minipage} \qquad
\caption{Construction of a PEPS wave function from a tensor-network operator: (a) We apply the TNO to a single-spin state $\ket{\psi}$; (b) this is interpreted as a PEPS tensor with four virtual legs, and a leg corresponding to the physical degrees of freedom; (c) This PEPS tensor is used to construct a many-body wave function in the thermodynamic limit.}
\label{fig:peps}
\end{figure}
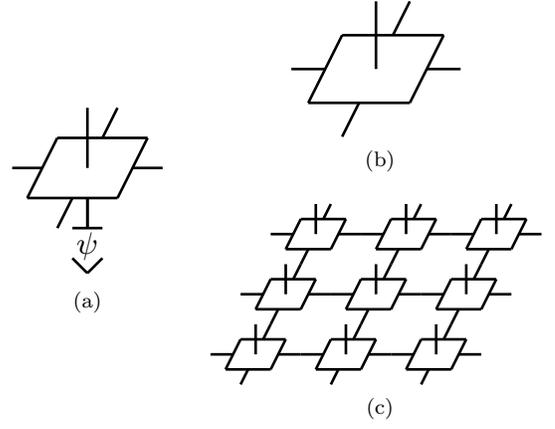

In this section we explain how to build a physical wave function from a TNO, and how we can apply clusters of local operators in an extensive way. Consider thereto as the elementary building block the object $T_{i_1,i_2,i_3,i_4}$ which represents an operator acting on a physical spin for every input of the virtual indices $\{i_1,i_2,i_3,i_4\}=0,\dots,D-1$ (here, $D$ is the dimension of the virtual legs and is called the bond dimension of the TNO). Alternatively, we can represent $T$ as a six-leg tensor,
\begin{equation}
T_{i_1,i_2,i_3,i_4} \quad\to\quad \tikzTNO{i_1}{i_2}{i_3}{i_4},
\end{equation}
where the up and down legs represent the action on a physical spin, and the four virtual legs correspond to the virtual indices. In Fig.~\ref{fig:peps} we have indicated how this tensor $T$ gives rise to a tensor-network operator $O(T)$ and, when applied to a product state, gives rise to a wave function on an infinite two-dimensional lattice.
\par Let us now show how to encode clusters of local operators $Q$ in this tensor. First of all, we can define the following entry
\begin{equation}
\tikzTNO{0}{0}{0}{0} \to \one + \beta Q,
\end{equation}
such that the corresponding TNO can be expanded in $\beta$ as
\begin{equation}
O(T) = \one + \beta \sum_i Q_i + \beta^2 \sum_{i\neq j} Q_i Q_j + \mathcal{O}(\beta^3).
\end{equation}
We can build two-site clusters of $Q$'s by defining a new entry
\begin{equation} \begin{array}{ll}
\tikzTNO{1}{0}{0}{0} \to \gamma_1 Q, & \tikzTNO{0}{1}{0}{0} \to  \gamma_1 Q,\\
\tikzTNO{0}{0}{1}{0} \to \gamma_1 Q, & \tikzTNO{0}{0}{0}{1} \to \gamma_1 Q,
\end{array} \end{equation}
where these four entries are related by rotations of the tensor. The corresponding expansion of the TNO in $\gamma_1$ is given by
\begin{multline}
O(T) = \one + \gamma_1 \sum_{\braket{ij}} Q_iQ_j \\ + \gamma_1^2 \sum_{(\braket{ij},\braket{kl})_d} Q_iQ_jQ_kQ_l + \mathcal{O}(\gamma_1^3).
\end{multline}
where $\braket{ij}$ denotes all nearest-neighbour pairs, and $(\braket{ij},\braket{kl})_d$ denotes all pairs of nearest-neighbours that do not overlap. Three-site clusters can be represented by the entries
\begin{align}
& \tikzTNO{1}{1}{0}{0} \to \gamma_2 \tilde{Q} \qquad (\text{and all rotations}),\\
& \tikzTNO{1}{0}{1}{0} \to \gamma_3 \tilde{Q} \qquad (\text{and all rotations}),
\end{align}
with corresponding expansions in the TNO:
\begin{multline}
O(T) = \one + \gamma_2\gamma_1^2 \sum_{(ijk)_l} O_i\tilde{Q}_jO_k \\ + \gamma_3\gamma_1^2 \sum_{(ijk)_c} O_i\tilde{Q}_jO_k + \mathcal{O}(\gamma_1^2\gamma_2^2) + \mathcal{O}(\gamma_1^2\gamma_3^2).
\end{multline}
Here we have used $(ijk)_l$ and $(ijk)_c$ for denoting all linear and corner-shaped three-site clusters, resp. The center site of each cluster is labeled $j$, and can correspond to a different operator $\tilde{Q}$.
\par One can imagine how larger and larger clusters can be constructed in this way. Also, by opening up more levels in the virtual indices, we can incorporate clusters of different operators in the TNO.
\par Note that we are never concerned with the normalization of the TNO; this is easily taken into account when optimizing the ground-state density expectation value \cite{Vanderstraeten16}.

\section{Transverse-field Ising model}

In this section we treat the transverse-field Ising model on the square lattice,
\begin{equation}
H_{\text{TFIM}} = - \lambda_0 \sum_i X_i - \lambda_1 \sum_{\braket{ij}} Z_iZ_j ,
\end{equation}
where $\braket{ij}$ denotes nearest-neighbour pairs of spins.

\subsection{Mean-field theory}

In mean-field theory we use the one-parameter family of states
\begin{equation}
\ket{\psi_\alpha} = \prod_{i} \left(1+\alpha Z_i \right) \ket{+X},
\end{equation}
with a magnetization
\begin{equation}
m(\alpha) = \frac{1}{N} \frac{\bra{\psi_\alpha} Z \ket{\psi_\alpha}}{\braket{\psi_\alpha|\psi_\alpha}} = \frac{2\alpha}{\alpha^2+1}.
\end{equation}
The energy density of this state is given by
\begin{equation} \begin{split}
e(\alpha) &= \frac{1}{N} \frac{\bra{\psi_\alpha} H_{\text{TFIM}} \ket{\psi_\alpha}}{\braket{\psi_\alpha|\psi_\alpha}},\\
&= - \lambda_0 \frac{1-\alpha^2}{1+\alpha^2} - \lambda_1 \frac{8\alpha^2}{(1+\alpha^2)^2}.
\end{split} \end{equation}
This function can be minimized straightforwardly for a given value of $\lambda_1/\lambda_0$, showing that the minimum is at $\alpha=0$ for $\lambda_1/\lambda_0<4$, and the minimum starts to shift once $\lambda_1/\lambda_0>4$. The resulting magnetization curve was plotted in the main body.

\subsection{Exponentiated perturbation theory}

In this section, we show explicitly how the exponentiated forms of perturbation theory reproduce the linear perturbative expansions, but can go beyond them as well. We treat both limits separately.

\noindent\emph{Around paramagnetic state}--- %
In this limit we start from $H_1=-\sum_{\braket{ij}} Z_iZ_j$ and treat $H_0=-\sum_iX_i$ as a perturbation with small prefactor $\lambda_0$; we will start from the state $\ket{\psi_1}$ and do perturbation theory up to second order in $\lambda_0$. At first order, we find $E_1=\bra{\psi_1} X_i \ket{\psi_1}=0$ and the following wave function:
\begin{align}
\ket{\psi} = \left( \one + \frac{\lambda_0}{8\lambda_1} \sum_i X_i \right) \ket{\psi_1}.
\end{align}
We can exponentiate this form to obtain
\begin{align}
\ket{\phi} &= \exp\left(\frac{\lambda_0}{8\lambda_1} \sum_i X_i \right) \ket{\psi_1} ,\\
& \sim \left( \one + \frac{\lambda_0}{8\lambda_1} \sum_i X_i  \right. \nonumber \\
& \hspace{1.5cm} \left. + \frac{\lambda_0^2}{64\lambda_1^2} \sum_{(ij)} X_i X_j + \mathcal{O}(\lambda_0^3) \right) \ket{\psi_1},
\end{align}
where we have denoted $\sum_{(ij)}$ as a sum over all pairs of sites, not necessarily nearest neighbour, and $i\neq j$. Note that each pair $\sum_{(ij)}$ is summed over only once in the above expression, explaining why the factor $\frac{1}{2}$ from the expansion of the exponential has vanished. In second order, we also have a contribution where two $X$'s act on the same state, giving rise to a component along the unperturbed state $\ket{\psi_1}$. Since these contributions are unimportant from the variational point of view, we have omitted them from the expansion. We observe that the exponentiated form not only contains the first-order expansion, but represents large portions of higher orders in $\lambda_0$ as well. Out of the region of small $\lambda_0$, we expect that this state contains more correlations than the linear form above, and therefore is a better variational state.
\par Let us go to second order. The second-order wave function is given by
\begin{multline} \label{eq:aroundparamagnetic2}
\ket{\psi} = \left( \one + \frac{\lambda_0}{8\lambda_1} \sum_i X_i + \frac{\lambda_0^2}{64\lambda_1^2} \sum_{(ij)_d}  X_iX_j \right. \\
\left. + \frac{\lambda_0^2}{48\lambda_1^2} \sum_{(ij)_n} X_i X_j \right) \ket{\psi_1} .
\end{multline}
Here, ${(ij)_n}$ represents all nearest-neighbour pairs, whereas ${(ij)_d}$ stands for all pairs of sites that are not on neighbouring sites. Upon exponentiating, we observe that the disconnected contributions (where $i$ and $j$ are not nearest neighbours) are already contained within the first-order wave function, but we need to correct the prefactor for the last term. Therefore, we have the following exponentiated version of the second-order wave function
\begin{multline}
\ket{\phi} = \exp\left(  \frac{\lambda_0^2}{192\lambda_1^2} \sum_{(ij)_n} X_iX_j \right) \\
\times \exp\left( \frac{\lambda_0}{8\lambda_1} \sum_i X_i \right) \ket{\psi_1},
\end{multline}
which again matches the above linear form up to second order, but contains a lot more than that.

\noindent\emph{Around polarized state}--- %
Now we start from \mbox{$H_0=-\sum_iX_i$} and treat $H_1=-\sum_{\braket{ij}}Z_iZ_j$ as a perturbation with small prefactor $\lambda_1$; we will start from the state $\ket{\psi_0}$, and derive the perturbation theory in $\lambda_1$. At first order, we find the following wave function:
\begin{align}
\ket{\psi} = \left( \one + \frac{\lambda_1}{4\lambda_0} \sum_{\braket{ij}} Z_iZ_j \right) \ket{\psi_0},
\end{align}
which we can exponentiate to obtain
\begin{align}
\ket{\phi} &= \exp\left(\sum_{\braket{ij}} \frac{\lambda_1}{4\lambda_0} Z_iZ_j \right) \ket{\psi_0}, \\
& \sim \left( \one + \frac{\lambda_1}{4\lambda_0} \sum_{\braket{ij}} Z_iZ_j + \frac{\lambda_1^2}{16\lambda_0^2} \sum_{(ijk)_c} Z_iZ_j^2 Z_l \right. \nonumber \\
& \left. \qquad  + \frac{\lambda_1^2}{16\lambda_0^2} \sum_{(\braket{ij}\braket{kl})_d} Z_iZ_j Z_kZ_l + \mathcal{O}(\lambda_1^3) \right) \ket{\psi_0}.
\end{align}
Here $(\braket{kl}\braket{ij})_d$ denotes all two nearest-neighbour pairs that do not have any sites in common, whereas we have used $(ijk)_c$ as a notation for all three-site clusters ($j$ denotes the center site of the cluster and the cluster can take on two different shapes).
\par The second-order result for the wave function is given by
\begin{multline} \label{eq:aroundpolarized2}
\ket{\psi} =  \left( \one + \frac{\lambda_1}{4\lambda_0} \sum_{\braket{ij}} Z_iZ_j + \frac{\lambda_1^2}{16\lambda_0^2} \sum_{(\braket{ij},\braket{kl})_d} Z_iZ_j Z_kZ_l \right. \\
\left. + \frac{\lambda_1^2}{8\lambda_0^2} \sum_{(ijk)_c} Z_iZ_j^2Z_k \right) \ket{\psi_0}.
\end{multline}
Again, we observe that the disconnected contributions are already contained within the first-order exponentiated wave function, but we need a correction for the second term. Therefore, we have the following exponentiated version of the second-order wave function
\begin{multline}
\ket{\phi} = \exp\left( \frac{\lambda_1^2}{16\lambda_0^2} \sum_{(ijk)_c} Z_iZ_j^2Z_k \right) \\ \times \exp\left(\frac{\lambda_1}{4\lambda_0} \sum_{\braket{ij}} Z_iZ_j \right) \ket{\psi_0}.
\end{multline}
This wave function matches the above linear form up to second order -- again, up to the components along the unperturbed state -- but contains a lot more.

\subsection{The TNO for the Ising model}

In this section we explain how to represent perturbative expansions for the TFIM using TNOs. As explained above, we define a TNO $O(T)$ by a single six-leg tensor $T$, represented diagrammatically as
\begin{equation}
T = \tikzTNO{i_1}{i_2}{i_3}{i_4}\;.
\end{equation}
The four virtual indices $\{i_1,i_2,i_3,i_4\}$ can take on values $0,1,2$ (i.e., the TNO has bond dimension $D=3$), whereas the up and down going legs correspond to the physical action of the tensor on a single-site spin state. The tensor thus represents a one-site operator for every specific input on the virtual level. In Table \ref{tab:TNO_ising} we list all the different non-zero entries that we possibly need for representing the series expansions in $\lambda_0$ and $\lambda_1$. Note in advance that the TNOs are \emph{not} exactly the same operators as the exponentiated operators that we have introduced above, but share the same ``extensive'' properties. Let us follow the TNO construction in detail.

\begin{table} \centering
\begin{tabular}{|c|c|}
\hline
\multicolumn{2}{|c|}{$\quad \tikzTNO{0}{0}{0}{0}\rightarrow \one + \beta X \quad$} \\
\hline
$\quad \tikzTNO{1}{0}{0}{0} \rightarrow \gamma_1 X    \quad$ & $\quad \tikzTNO{2}{0}{0}{0}\rightarrow \delta_1 Z \quad$\\
\hline
$\quad \tikzTNO{1}{1}{0}{0} \rightarrow \gamma_2 X \quad$ & $\quad \tikzTNO{2}{2}{0}{0}\rightarrow \delta_2 \one \quad$\\
\hline
$\quad \tikzTNO{1}{0}{1}{0} \rightarrow \gamma_3 X \quad$ & $\quad \tikzTNO{2}{0}{2}{0}\rightarrow \delta_3 \one \quad$\\
\hline
$\quad \tikzTNO{1}{1}{1}{0} \rightarrow \gamma_4 X    \quad$ & $\quad \tikzTNO{2}{2}{2}{0}\rightarrow \delta_4 Z \quad$\\
\hline
$\quad \tikzTNO{1}{1}{1}{1} \rightarrow \gamma_5 X \quad$ & $\quad \tikzTNO{2}{2}{2}{2}\rightarrow \delta_5 \one \quad$\\
\hline
\end{tabular}
\caption{The tensor-network operator for the Ising model, determined by the eleven parameters $\{\beta,\gamma_1\dots\gamma_5,\delta_1\dots\delta_5\}$. The tensor should still be symmetrized, in the sense that the same operators are asigned to the rotated versions of the above entries. If all parameters are zero, the TNO acts as the unit operator. The two-site $ZZ$ operators are captured by the $\delta_1$ coefficient. The tensor entries with coefficients $\delta_2$ and $\delta_3$ capture the center site of a three-site cluster of $ZZ^2 Z$ operations, whereas $\delta_1$ again captures the end-points -- $\delta_2$ corresponds L-shaped clusters, whereas $\delta_3$ makes linear ones. Note that we can also make larger clusters by including more contributions from $\delta_2$ and $\delta_3$; the $\delta_4$ and $\delta_5$ entries also make larger and larger clusters. Similarly, on the $X$ side, we can make clusters of $X$ operators of increasing size.}
\label{tab:TNO_ising}
\end{table}

\noindent\emph{Around paramagnetic state}--- %
The first step in representing the perturbed state \eqref{eq:aroundparamagnetic2} is rather trivial. Indeed, by turning on the parameter $\beta$ in Table \ref{tab:TNO_ising} we can easily represent the appearance of local $X$ operations acting on the state $\ket{\psi_1}$. The two-site clusters of $XX$ operations can now be implemented by the $\gamma_1$ parameter. The expansion of the TNO with these two parameters is given by
\begin{multline}
O(T) = \one + \beta \sum_i X_i + \beta^2 \sum_{(ij)}  X_iX_j  \\
 + \gamma_1^2 \sum_{(ij)_n} X_i X_j + \mathcal{O}(\beta^3) + \mathcal{O}(\gamma_1^4) .
\end{multline}
This implies that the choice
\begin{equation}
\beta=\frac{\lambda_0}{8\lambda_1}, \quad \gamma_1=\sqrt{\frac{\lambda_0^2}{192\lambda_1^2}},
\end{equation}
reproduces the linear perturbative expansion up to second order. The other parameters $\gamma_2\dots\gamma_5$ can make larger clusters of $X$'s, corresponding to higher orders of perturbation theory.

\noindent\emph{Around polarized state}--- %
Now we will also need to include three-site clusters in order to reproduce second-order perturbation theory in $\lambda_1$. These larger clusters are provided by the $\delta_2$ and $\delta_3$ parameters in Table \ref{tab:TNO_ising}. The former represents the midpoint of a corner-shaped three-site cluster, whereas the latter is responsible for making the linear clusters. In both cases the tensor entry with parameter $\delta_1$ provides the end points of the cluster. This implies that the expansion of the TNO with these three parameters is given by
\begin{align}
O(T) = \one  &+ \delta_1^2 \sum_{(ij)_n} Z_iZ_j + \delta_1^4 \sum_{(\braket{ij},\braket{kl})_d} Z_iZ_j Z_kZ_l \nonumber \\
& + \delta_1^2\delta_2 \sum_{(ijk)_c} Z_iZ_k  +  \delta_1^2\delta_3 \sum_{(ijk)_l} Z_iZ_k \\
& + \mathcal{O}(\delta_1^6) + \mathcal{O}(\delta_1^2\delta_2^2+\delta_1^2\delta_3^2) \nonumber .
\end{align}
Here, $(ijk)_c$ and $(ijk)_l$ are corner-shaped and linear three-site clusters, resp. If we now choose the parameters
\begin{equation}
\delta_1 = \sqrt{\frac{\lambda_1}{4\lambda_0}}, \quad \delta_2=\delta_3=\frac{\lambda_1}{2\lambda_0},
\end{equation}
we reproduce the linear wave function \eqref{eq:aroundpolarized2} up to second order in $\lambda_1/\lambda_0$.

\noindent\emph{Variational parameters}--- %
The above then shows the explicit form of the TNOs that we have considered in order to produce the results in the main text. The mean-field result is obtained with $\alpha$ the only parameter; the first-order result is obtained by keeping $\{\alpha,\beta,\delta_1\}$ as variational parameters; the second-order result by keeping $\{\alpha,\beta,\delta_1,\gamma_1,\delta_2=\delta_3\}$.

\section{The Toric Code in magnetic field}

The Hamiltonian is given by
\begin{equation}
H_{\text{TCX}}= - \lambda_0 \sum_i X_i -\lambda_1 \left( \sum_v A_v + \sum_p B_p \right) ,
\end{equation}
where
\begin{equation}
A_v \prod_{j \in v}X_j, \quad
B_p= \prod_{j \in p}Z_j.
\end{equation}
We work in the subspace where $\braket{A_v}$ is always $1$, so we don't consider this part of the Hamiltonian in the following.

\subsection{Mean-field theory and virtual order parameter}

In order to reproduce mean-field theory for the toric code we introduce the one-parameter family of states
\begin{equation}
\ket{\psi_\alpha} = \prod_p \left( 1+\alpha B_p \right) \prod_i \ket{+X},
\end{equation}
For $\alpha=0$, we have the ground state of the $\lambda_1=0$ model, whereas the toric-code state is obtained for $\alpha=1$. The calculations for the variational optimization of this one-parameter ansatz can be found in Ref.~\cite{Dusuel15}.

\begin{figure}
\begin{equation*}
\begin{tikzpicture}[baseline = (X.base),every node/.style={scale=1},scale=.45]
\drawone (-5,0) -- (5,0); \drawone (-5,4) -- (5,4); \drawone (-5,-4) -- (5,-4);
\drawone (0,-5) -- (0,5); \drawone (4,-5) -- (4,5); \drawone (-4,-5) -- (-4,5);
\drawdot{-4}{+2} \drawdot{0}{+2} \drawdot{4}{+2}
\drawdot{-4}{-2} \drawdot{0}{-2} \drawdot{4}{-2}
\drawdot{+2}{-4} \drawdot{+2}{0} \drawdot{+2}{+4}
\drawdot{-2}{-4} \drawdot{-2}{0} \drawdot{-2}{+4}
\drawred (-3,0) -- (0,-3); \drawred (0,-3) -- (3,0); \drawred (3,0) -- (0,3); \drawred (0,3) -- (-3,0);
\drawred (4,1) -- (1,4); \drawred (1,4) -- (2,5); \drawred (4,1) -- (5,2);
\drawred (-4,1) -- (-1,4); \drawred (-1,4) -- (-2,5); \drawred (-4,1) -- (-5,2);
\drawred (4,-1) -- (1,-4); \drawred (1,-4) -- (2,-5); \drawred (4,-1) -- (5,-2);
\drawred (-4,-1) -- (-1,-4); \drawred (-1,-4) -- (-2,-5); \drawred (-4,-1) -- (-5,-2);
\end{tikzpicture}
\end{equation*}
\caption{The blocking of four spins on every second vertex into one supersite. To each supersite, a tensor of a TNO is associated. In this blocking configuration, the plaquette operators couple nearest-neighbour supersites by acting on two spins of each nearest-neighbour pair.}
\label{fig:tc_blocking}
\end{figure}
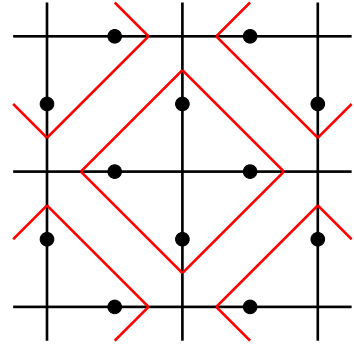

\par This reference state is produced by a TNO defined by the tensor $T$
\begin{equation}
T_{i_1,i_2,i_3,i_4} = \tikzTNOt{i_1}{i_2}{i_3}{i_4}.
\end{equation}
We have grouped four spins $\{a,b,c,d\}$ on every second vertex into one big supersite (see Fig.~\ref{fig:tc_blocking}), so that the plaquette operators always act on two spins of each supersite. As before, the action of the TNO tensor $T$ depends on the inputs of the four virtual indices $\{i_1,i_2,i_3,i_4\}$ which can take on two different values $0,1$. The different entries of the tensor are listed in Table \ref{tab:TNO_toric}. If we choose the values
\begin{equation} \begin{split}
& \alpha_1=\alpha^{1/2}, \quad  \alpha_2=\alpha_3=\alpha, \\ & \alpha_4=\alpha^{3/2}, \quad \alpha_5=\alpha^2,
\end{split} \end{equation}
the TNO is exactly equal to the operator that appears in $\ket{\psi_\alpha}$.
\par Also, this tensor contains the virtual order parameter. Indeed, from the framework of $\mathsf{G}$-injectivity we know that this TNO can give rise to a topologically ordered state only if the tensor $T$ satisfies the condition
\begin{multline}
 \sum_{i_1',i_2',i_3',i_4'} X_{i_1,i_1'} X_{i_2,i_2'} X_{i_3,i_3'} X_{i_4,i_4'} T_{i_1',i_2',i_3',i_4'} \\ = T_{i_1,i_2,i_3,i_4},
\end{multline}
i.e., acting with $X$'s on the virtual indices should leave the tensor invariant ($X_{i,i'}$ is the usual Pauli matrix). This is the non-trivial $\mathbb{Z}_2$-symmetry operation that is discussed in the main text. We define the virtual order parameter $v$ as
\begin{multline}
v = \left\lVert \; T_{i_1,i_2,i_3,i_4}   \phantom{\sum_{i_1'}} \right. \\
\left. - \sum_{i_1',i_2',i_3',i_4'} X_{i_1,i_1'} X_{i_2,i_2'} X_{i_3,i_3'} X_{i_4,i_4'} T_{i_1',i_2',i_3',i_4'} \; \right\rVert,
\end{multline}
where $\left\lVert \dots \right\rVert$ of a tensor just denotes the square root of the sum of all tensor entries squared. It can only be zero under the conditions
\begin{equation}
\alpha_1=\alpha_4,\qquad \alpha_5=1,
\end{equation}
which implies that the mean-field state $\ket{\psi_\alpha}$ only exhibits topological order if $\alpha=1$.

\begin{table} \centering
\begin{tabular}{|c|c|}
\hline
$\quad \tikzTNOt{0}{0}{0}{0} \rightarrow \one_a \otimes \one_b \otimes \one_c \otimes \one_d \quad$ \\
\hline
$\quad \tikzTNOt{0}{0}{0}{1} \rightarrow \alpha_1 Z_a \otimes Z_b \otimes \one_c \otimes \one_d \quad$ \\
\hline
$\quad \tikzTNOt{0}{0}{1}{1} \rightarrow \alpha_2 Z_a \otimes \one_b \otimes \one_c \otimes Z_d \quad$ \\
\hline
$\quad \tikzTNOt{0}{1}{0}{1} \rightarrow \alpha_3 Z_a \otimes Z_b \otimes Z_c \otimes Z_d \quad$ \\
\hline
$\quad \tikzTNOt{0}{1}{1}{1} \rightarrow \alpha_4 Z_a \otimes \one \otimes Z_c \otimes \one_d \quad$ \\
\hline
$\quad \tikzTNOt{1}{1}{1}{1} \rightarrow \alpha_5 \one_a \otimes \one_b \otimes \one_c \otimes \one_d \quad$ \\
\hline
\end{tabular}
\caption{The tensor-network operator for building the reference state $\ket{\psi_\alpha}$. Again, the tensor should still be symmetrized, in the sense that the same operators are asigned to the rotated versions of the above entries.}
\label{tab:alpha_toric}
\end{table}

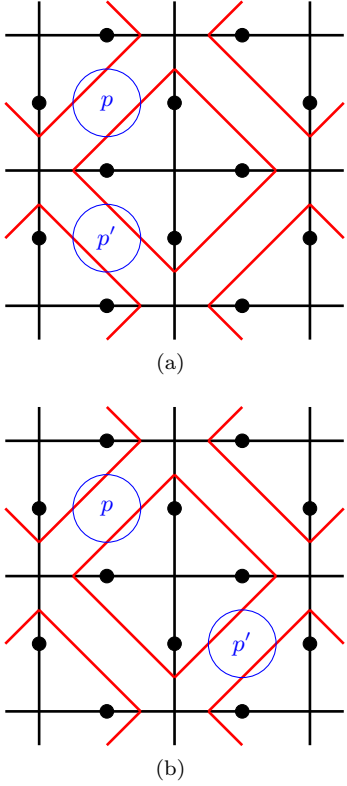
\begin{figure} \centering
\subfloat[]{
$\begin{tikzpicture}[baseline = (X.base),every node/.style={scale=1},scale=.45]
\drawone (-5,0) -- (5,0); \drawone (-5,4) -- (5,4); \drawone (-5,-4) -- (5,-4);
\drawone (0,-5) -- (0,5); \drawone (4,-5) -- (4,5); \drawone (-4,-5) -- (-4,5);
\drawdot{-4}{+2} \drawdot{0}{+2} \drawdot{4}{+2}
\drawdot{-4}{-2} \drawdot{0}{-2} \drawdot{4}{-2}
\drawdot{+2}{-4} \drawdot{+2}{0} \drawdot{+2}{+4}
\drawdot{-2}{-4} \drawdot{-2}{0} \drawdot{-2}{+4}
\drawred (-3,0) -- (0,-3); \drawred (0,-3) -- (3,0); \drawred (3,0) -- (0,3); \drawred (0,3) -- (-3,0);
\drawred (4,1) -- (1,4); \drawred (1,4) -- (2,5); \drawred (4,1) -- (5,2);
\drawred (-4,1) -- (-1,4); \drawred (-1,4) -- (-2,5); \drawred (-4,1) -- (-5,2);
\drawred (4,-1) -- (1,-4); \drawred (1,-4) -- (2,-5); \drawred (4,-1) -- (5,-2);
\drawred (-4,-1) -- (-1,-4); \drawred (-1,-4) -- (-2,-5); \drawred (-4,-1) -- (-5,-2);
\draw[color=blue] (-2,2) circle (1); \draw[color=blue] (-2,2) node {$p$};
\draw[color=blue] (-2,-2) circle (1); \draw[color=blue] (-2,-2) node {${p'}$};
\end{tikzpicture}$} \qquad
\subfloat[]{
$\begin{tikzpicture}[baseline = (X.base),every node/.style={scale=1},scale=.45]
\drawone (-5,0) -- (5,0); \drawone (-5,4) -- (5,4); \drawone (-5,-4) -- (5,-4);
\drawone (0,-5) -- (0,5); \drawone (4,-5) -- (4,5); \drawone (-4,-5) -- (-4,5);
\drawdot{-4}{+2} \drawdot{0}{+2} \drawdot{4}{+2}
\drawdot{-4}{-2} \drawdot{0}{-2} \drawdot{4}{-2}
\drawdot{+2}{-4} \drawdot{+2}{0} \drawdot{+2}{+4}
\drawdot{-2}{-4} \drawdot{-2}{0} \drawdot{-2}{+4}
\drawred (-3,0) -- (0,-3); \drawred (0,-3) -- (3,0); \drawred (3,0) -- (0,3); \drawred (0,3) -- (-3,0);
\drawred (4,1) -- (1,4); \drawred (1,4) -- (2,5); \drawred (4,1) -- (5,2);
\drawred (-4,1) -- (-1,4); \drawred (-1,4) -- (-2,5); \drawred (-4,1) -- (-5,2);
\drawred (4,-1) -- (1,-4); \drawred (1,-4) -- (2,-5); \drawred (4,-1) -- (5,-2);
\drawred (-4,-1) -- (-1,-4); \drawred (-1,-4) -- (-2,-5); \drawred (-4,-1) -- (-5,-2);
\draw[color=blue] (-2,2) circle (1); \draw[color=blue] (-2,2) node {$p$};
\draw[color=blue] (2,-2) circle (1); \draw[color=blue] (2,-2) node {${p'}$};
\end{tikzpicture}$}
\caption{Different configurations of plaquette pairs $p$ and ${p'}$. On the top we have two plaquette operators that overlap on one site, which we denote by $(pp')_o$. On the bottom we have two neighbouring plaquette operators that do not have a spin in common, which we denote by $(pp')_n$. All other configurations, where the two plaquettes $p$ and $p'$ are further away, are denoted by $(pp')_d$.}
\label{fig:plaquettes}
\end{figure}

\subsection{The TNO for the Toric Code}

We can capture the two second-order perturbative expansions with the TNO of bond dimension $D=3$ that is given in Table \ref{tab:TNO_toric}. Let us follow the construction explicitly in both limiting cases.

\begin{table*} \centering
\begin{tabular}{|p{8cm}|p{8cm}|}
\hline
\multicolumn{2}{|p{16cm}|}{$\begin{aligned}[c]
 \quad \tikzTNOt{0}{0}{0}{0} \rightarrow \one_a \otimes \one_b \otimes \one_c \otimes \one_d
 + \beta_1 \left\{ \begin{array}{c} X_a \otimes \one_b \otimes \one_c \otimes \one_d \\ \one_a \otimes X_b \otimes \one_c \otimes \one_d \\ \one_a \otimes \one_b \otimes X_c \otimes \one_d \\ \one_a \otimes \one_b \otimes \one_c \otimes X_d \end{array} \right\}
 + \beta_2 \left\{ \begin{array}{c} X_a \otimes X_b \otimes \one_c \otimes \one_d \\ \one_a \otimes X_b \otimes X_c \otimes \one_d \\ \one_a \otimes \one_b \otimes X_c \otimes X_d \\ X_a \otimes \one_b \otimes \one_c \otimes X_d \end{array} \right\} \end{aligned}$ }\\
\hline
$ \quad \tikzTNOt{0}{0}{0}{1} \rightarrow \gamma_1 O_{ab} \quad$ &
$\quad \tikzTNOt{0}{0}{1}{1} \rightarrow \gamma_2 O_{ab} O_{bc} \quad$ \\
\hline
$\quad \tikzTNOt{0}{0}{0}{2} \rightarrow \delta_1 Z_a \otimes Z_b \otimes \one_c \otimes \one_d \quad$ &
$\quad \tikzTNOt{0}{0}{2}{2} \rightarrow \delta_2 Z_a \otimes \one_b \otimes Z_c \otimes \one_d \quad$ \\
\hline
\end{tabular}
\caption{The second tensor-network operator for the TCXM. The two-spin operator that we have used is given by the prescription $O_{ab}=(X_a\otimes \one_b + \one_a \otimes X_b ) \otimes \one_c \otimes \one_d$, and the same for $O_{bc}$, $O_{cd}$ and $O_{da}$.}
\label{tab:TNO_toric}
\end{table*}

\noindent\emph{Around toric-code state}--- %
Suppose now that we start from the toric-code state $\ket{\psi_1}$, which is the ground state of $H_1=-\lambda_1\sum_pB_p$, and we have $V=-\sum_iX_i$. In second-order perturbation theory in $\lambda_0$ we find
\begin{multline}
\ket{\psi} = \Bigg( \one + \frac{\lambda_0}{4\lambda_1} \sum_i X_i + \frac{\lambda_0^2}{16\lambda_1^2} \sum_{(ij)_d} X_i X_j \\ + \frac{\lambda_0^2}{8\lambda_1^2} \sum_{(ij)_p} X_i X_j \Bigg) \ket{\psi_1},
\end{multline}
where $(ij)_d$ represents two sites that are not located on the same plaquette, and $(ij)_p$ are two spins that share a plaquette. In order to represent this state with the TNO we turn on the three parameters $\beta_1$, $\beta_2$ and $\delta_1$. Indeed, the expansion in terms of these three parameters is given by
\begin{align}
O(T) = \one &+ \beta_1 \sum_i X_i + \beta_1^2 \sum_{(ij)_d} X_iX_j + \mathcal{O}(\beta_1^3) \nonumber \\
&+ \beta_2 \sum_{(ij)_s} X_iX_j + \mathcal{O}(\beta_2^2) \nonumber \\
&+ \delta_1^2 \sum_{(ij)_p} X_i X_j + \mathcal{O}(\delta_1^4),
\end{align}
where ${(ij)_d}$ represent pairs of spin that do not live on the same supersite, $(ij)_s$ represent pairs of spins that do live on the same supersite, and $(ij)_p$ represent pairs of spin that live on the same plaquette. Therefore, the choice
\begin{equation}
\beta_1=\frac{\lambda_0}{4\lambda_1}, \quad \beta_2=\frac{\lambda_0^2}{16\lambda_1^2}, \quad \delta_1=\sqrt\frac{\lambda_0^2}{8\lambda_1^2},
\end{equation}
recovers the linear perturbed state up to second order in $\lambda_0$.

\par\noindent\emph{Around polarized state}--- %
Suppose now that we start from the fully polarized state $\ket{\psi_{0}}$, which is the ground state of $H_0=-\lambda_0 \sum_iX_i$,  and we have $V=-\sum_pB_p$. In second order in perturbation theory in $\lambda_1$ we find
\begin{multline}
\ket{\psi} =  \left( \one + \frac{\lambda_1}{8\lambda_0}\sum_p B_p + \frac{\lambda_1^2}{64\lambda_0^2} \sum_{(pp')_d} B_p B_{p'} \right. \\ \left. + \frac{\lambda_1^2}{64\lambda_0^2} \sum_{(pp')_n} B_p B_{p'} + \frac{\lambda_1^2}{48\lambda_0^2} \sum_{(pp')_o} B_p B_{p'} \right) \ket{\psi_{0}},
\end{multline}
where $(pp')_n$ denotes a pair of neighbouring plaquettes (which don't share any spins), $(pp')_o$ denotes a pair of overlapping plaquettes (which share one common spin), and $(pp')_d$ denotes a pair of disconnected plaquettes (see Fig.~\ref{fig:plaquettes}). This state can be obtained with the TNO of Table~\ref{tab:TNO_toric} up to second order in $\lambda_1$. Indeed, the parameter $\gamma_1$ introduces end-points of clusters of plaquettes, whereas $\gamma_2$ introduces overlapping plaquette pairs $(pp')_o$, and $\gamma_3$ corresponds to neighbouring plaquette pairs $(pp')_n$ (see Fig.~\ref{fig:plaquettes}). This implies the following form of the TNO:
\begin{align}
O(T) = \one &+ \gamma_1^2 \sum_p B_p + \gamma_1^4 \sum_{(pp')_d} B_p B_{p'} \nonumber \\
&+ \gamma_1^2\gamma_2 \sum_{(pp')_n} B_p B_{p'} + \gamma_1^2\gamma_3 \sum_{(pp')_o} B_p B_{p'} \nonumber \\
&+ \mathcal{O}(\gamma_1^6) + \mathcal{O}(\gamma_1^2\gamma_2^2) + \mathcal{O}(\gamma_1^2\gamma_3^2).
\end{align}
If we now we fix the parameters as
\begin{equation}
\gamma_1 = \sqrt\frac{\lambda_1}{8\lambda_0}, \qquad \gamma_2=\frac{\lambda_1}{8\lambda_0}, \qquad \mu_3 = \frac{\lambda_1}{6\lambda_0},
\end{equation}
we recover the linear form of the perturbed state up to second order in $\lambda_1$.

\subsection{Stacking TNOs}

We can now construct a number of variational ansatz states by including, order by order, the parameters in the above TNO $O(T)$ and applying it to the reference state $\ket{\psi_\alpha}$. Since the reference state is itself a $D=2$ PEPS, the action with a $D=3$ TNO gives rise to a class of PEPS with bond dimension $D=6$. In order to reduce this bond dimension, we introduce a slightly different construction of the variational ansatz.
\par The essential modification is that we include the perturbative expansion around the polarized state in the same TNO [Table~\ref{tab:alpha_toric}] as the one we have used for constructing the reference state $\ket{\psi_\alpha}$. Then, in a next step, we can apply the TNO that we need to represent the expansion around the toric-code state.
\begin{widetext}
The variational ansatz that we have used to obtain the results as reported in the main body, is given by
\begin{equation}
\ket{\beta_1\dots\beta_5,\gamma_1\dots\gamma_5,\delta_1\dots\delta_5} = \\ O(T^1_{\{\beta_1\dots\beta_5,\gamma_1\dots\gamma_5\}}) O(T^2_{\{\delta_1\dots\delta_5\}}) \prod_{i}\ket{+},
\end{equation}
where in the definition of the tensor $T^1$ we have introduced the following operator on a supersite:
\begin{multline}
\hspace{1cm} P^X_{abcd} = \one_a \otimes \one_b \otimes \one_c \otimes \one_d
 + \beta_1 \left\{ \begin{array}{c} X_a \otimes \one_b \otimes \one_c \otimes \one_d \\ \one_a \otimes X_b \otimes \one_c \otimes \one_d \\ \one_a \otimes \one_b \otimes X_c \otimes \one_d \\ \one_a \otimes \one_b \otimes \one_c \otimes X_d \end{array} \right\}
 + \beta_2 \left\{ \begin{array}{c} X_a \otimes X_b \otimes \one_c \otimes \one_d \\ \one_a \otimes X_b \otimes X_c \otimes \one_d \\ \one_a \otimes \one_b \otimes X_c \otimes X_d \\ X_a \otimes \one_b \otimes \one_c \otimes X_d \end{array} \right\} \\
 + \beta_3 \left\{ \begin{array}{c} X_a \otimes \one_b \otimes X_c \otimes \one_d \\ \one_a \otimes X_b \otimes \one_c \otimes X_d  \end{array} \right\}
 + \beta_4 \left\{ \begin{array}{c} X_a \otimes X_b \otimes X_c \otimes \one_d \\ \one_a \otimes X_b \otimes X_c \otimes X_d \\ X_a \otimes \one_b \otimes X_c \otimes X_d \\ X_a \otimes X_b \otimes \one_c \otimes X_d \end{array} \right\}
 + \beta_5 X_a \otimes X_b \otimes X_c \otimes X_d.  \hspace{1cm}
\end{multline}
\end{widetext}

\begin{table*} \centering
\begin{tabular}{|p{8cm}|p{8cm}|}
\hline
$ \quad \tikzTNOt{0}{0}{0}{0} \rightarrow P_{X,abcd}(\{\beta_1,\dots\beta_5\}) $ &
$ \quad \tikzTNOt{0}{0}{0}{0} \rightarrow \one_a \otimes \one_b \otimes \one_c \otimes \one_d $ \\
\hline
$\quad \tikzTNOt{0}{0}{0}{1} \rightarrow \gamma_1 O_{ab} \quad$ &
$\quad \tikzTNOt{0}{0}{0}{1} \rightarrow \delta_1 Z_a \otimes Z_b \otimes \one_c \otimes \one_d \quad$ \\
\hline
$\quad \tikzTNOt{0}{0}{1}{1} \rightarrow \gamma_2 O_{ab} O_{bc} \quad$ &
$\quad \tikzTNOt{0}{0}{1}{1} \rightarrow \delta_2 Z_a \otimes \one_b \otimes Z_c \otimes \one_d \quad$ \\
\hline
$\quad \tikzTNOt{0}{1}{0}{1} \rightarrow \gamma_3 O_{ab} O_{cd} \quad$ &
$\quad \tikzTNOt{0}{1}{0}{1} \rightarrow \delta_3 Z_a \otimes Z_b \otimes Z_c \otimes Z_d \quad$ \\
\hline
$\quad \tikzTNOt{0}{1}{1}{1} \rightarrow \gamma_4 O_{ab} O_{bc} O_{cd} \quad$ &
$\quad \tikzTNOt{0}{1}{1}{1} \rightarrow \delta_4 Z_a \otimes \one \otimes Z_c \otimes \one_d \quad$ \\
\hline
$\quad \tikzTNOt{1}{1}{1}{1} \rightarrow \gamma_5 O_{ab} O_{bc} O_{cd} O_{da} \quad$ &
$\quad \tikzTNOt{1}{1}{1}{1} \rightarrow \delta_5 \one_a \otimes \one_b \otimes \one_c \otimes \one_d \quad$\\
\hline
\end{tabular}
\caption{The two tensor-network operators $T^1_{\{\beta_1\dots\beta_5,\gamma_1\dots\gamma_5\}}$ (left column) and $T^2_{\{\gamma_1\dots\gamma_5\}}$ (right column).The two-spin operator that we have used is given by the prescription $O_{ab}=(X_a\otimes \one_b + \one_a \otimes X_b ) \otimes \one_c \otimes \one_d$, and the same for $O_{bc}$, $O_{cd}$ and $O_{da}$.}
\label{tab:TNO_toric2}
\end{table*}

\section{XXZ model}

In this section, we explain how to simulate the XXZ model on the square lattice with our variational ansatz. First of all, we perform a sublattice rotation of the original model in order to arrive at the Hamiltonian
\begin{equation}
H = - \sum_{\braket{ij}} Z_iZ_j + \lambda \sum_{\braket{ij}} \left( X_i X_j - Y_i Y_j \right).
\end{equation}
This rotation maps a staggered magnetization in the $Y$ and $Z$ direction into a uniform one. For $\lambda=0$ an exact ground state is given by
\begin{equation*}
\ket{\psi_0} = \ket{+Z},
\end{equation*}
whereas in first-order perturbation theory we have
\begin{equation}
\ket{\psi} = \left( \one - \lambda \frac{1-P_0}{H_0-E_0} \sum_{\braket{ij}} (X_iX_j - Y_iY_j) \right) \ket{\psi_0}.
\end{equation}
Since we know
\begin{align}
( X_iX_j - Y_iY_j ) \ket{+Z}_i \ket{+Z}_j = 2 \ket{-Z}_i \ket{-Z}_j,
\end{align}
the wave function can be simplified to yield
\begin{equation}
\ket{\psi} = \left( \one - \frac{\lambda}{6} \sum_{\braket{ij}} S^-_iS^-_j \right) \ket{\psi_0}.
\end{equation}
The TNO that gives rise to this wave function up to first order is given by a $D=2$ tensor given by the two entries
\begin{equation}
\tikzTNO{0}{0}{0}{0} \to \one, \quad \tikzTNO{1}{0}{0}{0} \to i\sqrt{\frac{\lambda}{6}} S^-,
\end{equation}
and all rotated versions of the second entry. By interpreting the weight of the second entry as a variational parameter, we obtain the results as reported in the main text. 
\par Note that the resulting wavefunction is $\mathrm{U}(1)$-invariant, since the tensor is invariant under the $\mathrm{U}(1)$ generators $s_z=\tfrac12(\ket0\bra0-\ket1\bra1)$ and $S_z=\ket1\bra1$ acting on the physical and virtual indices, respectively, while the initial N\'{e}el state provides a (staggered) $\mathrm{U}(1)$ background charge of $\tfrac12$ per site.

\end{document}